# Robustness of Spin Polarization in Graphene-based Spin Valves

by Masashi Shiraishi[*], Megumi Ohishi, Ryo Nouchi, Nobuhiko Mitoma, Takayuki Nozaki, Teruya Shinjo and Yoshishige Suzuki

[*] Prof. Masashi Shiraishi

Graduate School of Engineering Science, Osaka University,

1-3 Machikaneyama-cho, Toyonaka, Osaka 560-8531, Japan

and

PRESTO-JST, 4-1-8 Honcho, Kawaguchi 332-0012, Saitama, Japan

e-mail ) shiraishi@mp.es.osaka-u.ac.jp

Megumi Ohishi, Ryo Nouchi, Nobuhiko Mitoma, Takayuki Nozaki, Teruya Shinjo, Yoshishige Suzuki

Graduate School of Engineering Science, Osaka University,

1-3 Machikaneyama-cho, Toyonaka, Osaka 560-8531, Japan

Keywords : Spin injection, Spin current, Graphene, Spin polarization, Non-local

# Cover Poster

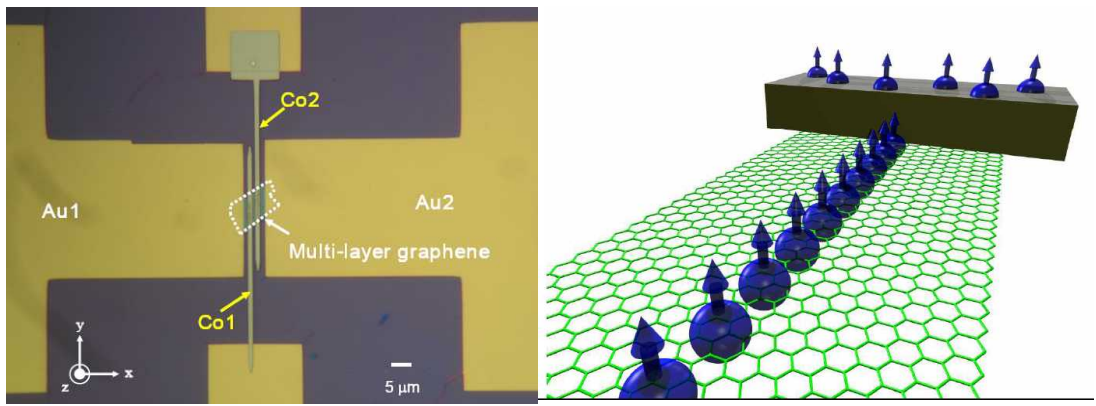

Decrease of spin polarization in spintronics devices under an application of bias voltage is one of currently important problems which should be solved. We reveal unprecedented robustness of spin polarization in multi-layer graphene spin valves at room temperature. Surprisingly, the spin polarization of injected spins is constant up to a bias voltage of +2.7 V and -0.6 V in positive and negative bias voltage applications at room temperature, which is superior to all spintronics devices. Our finding is induced by suppression of spin scattering due to an ideal interface formation. Furthermore, we find important accordance between theory and experiment in molecular spintronics by observing the fact that signal intensity in a local scheme is double of that in a non-local scheme as theory predicts, which provides construction of a steadfast physical basis in this field.

# 1. Introduction

Spin injection and transport phenomena give rise to magnetoresistance (MR) effects, and are investigated in the emerging research field, spintronics, where molecular materials including nano-carbonaceous molecules such as carbon nanotube and graphene in addition to inorganic metals and semiconductors are utilized[1-12]. The degree of MR effects is determined by the polarization of injected spins. The MR ratio decreases monotonically as the bias voltage increases, and $V_{half}$ is defined as the bias voltage when the MR ratio becomes a half of the maximum. For instance, $V_{half}$ is ca. +1 V in MgO-based TMR devices[6] at room temperature (RT), which is the best $V_{half}$ value among all spintronics devices. However, this value is not sufficient for designing various types of spin devices, and much effort has been paid for the improvement. Contrary to MgO-based TMR devices, $V_{half}$ values for single-walled carbon nanotube (SWNT)[13] and multi-layer-graphene (MLG) based[14] spin valves have been reported to be 10 and 5 mV, respectively, even below liquid helium temperatures, whereas multi-walled carbon nanotube (MWNT) exhibited comparably better bias voltage dependence at 5 K ($V_{half}$ = 200 mV)[15]. Recently, spin transport in amorphous rubrene($C_{42}H_{28}$)-based spin valves has been reported, however the $V_{half}$ was estimated to be still 100-200 mV at RT[16]. Such inferior bias voltage dependence of the MR ratio (= spin polarization of injected spins) is one of the major disadvantages of molecular spintronics. Hence, there has been a strong demand in molecular spintronics that $V_{half}$ should be comparable to that in MgO-based devices for further progress. Here, it is noteworthy that previous spin injection studies[13-16] were implemented by exploiting only the "local" scheme, and that the discrepancy between the results of non-local and local measurements has been pointed out[11,17], where the resistance hysteresis in the local scheme was abnormally large compared with that in the non-local scheme and this may be attributed to the fact that most of the local signal was not due to spin accumulation. Based on the backgrounds mentioned above, it is worthwhile to implement (1) precise clarification of the bias voltage dependence of spin polarization in molecular spin devices by comparing the results from both non-local and local schemes and (2) verification of the fact that the "local" signal intensity is double of the "non-local" signal as theory of an intensity of a spin injection signal in non-local and local schemes (Chapters II and III of ref. 17) predicts and as exhibited in the experiment in metallic spin valves in the same study[18]. In this study, we found an unprecedented robustness of spin polarization of injected spins in a graphene-based spin device, which was attributed to an ideal interface formation between ferromagnet and graphene. The good accordance between a theory and an experiment in molecular spintronics was verified by observing that signal intensity in a local scheme is double of that in a non-local scheme as the theory predicts, which corroborates the ideal interface formation and provides construction of a steadfast physical basis in this field. The robustness was more than 10 times greater than that in previous molecular spintronics studies[13-16], and it was also much superior to that in MgO-based TMR devices[6]. Graphene spintronics is currently expected to be a potential candidate for future spintronics, and the spin injection/precession and the observation of comparably long spin coherent length and time have been achieved at room temperature. However, the former had been also implemented in metallic spin devices, and the spin coherent time and length were still one to two orders of magnitude smaller than those in Si-based spin devices[19] and also better spin coherence have been reported in the MWNT spin valves at 5 K[15]. Recently, spin transport properties in single layer graphene (SLG) were reported[20,21], where electronic spin drift and anisotropic spin relaxation were investigated, and fundamental spin transport properties have been understood, such as importance of a Elliott-Yafet mechanism for spin scattering in SLG and a possibility of control of spin relaxation. These studies exhibited high potential of graphene spintronics. Although a spin injection signal intensity and spin coherence should be further improved compared with those in MWNT[15], our finding in this study provides a part of superiority of graphene spintronics compared with the other spintronics devices in the robustness of spin polarization of injected spins in spin valve devices, which can induce a further development not only in graphene spintronics but also in all of spintronics.

## 2. Results and discussion

Figure 2(a) shows non-local resistance in the MLG spin valve at RT. The non-local resistance is almost constant (~18 m$\Omega$) under various conditions of the injection current; in the other words, the output voltage, $\Delta V_{non-local}$, has linear dependence for the injected current within ±1 mA (see Fig. 2(b)). Here, $\Delta V_{non-local}$ is defined as ($V_P$-$V_{AP}$), where $V_P$ and $V_{AP}$ are the non-local voltages for the parallel and anti-parallel configurations of Co1 and Co2, respectively. An offset voltage was observed in addition to the clear resistance hysteresis. We think the origin of the offset voltage is a large width of the MLG channel. As shown in Fig. 1, the channel width in our sample was more than 7 $\mu$m, which was comparably larger than those in the other studies. In such a case, a voltage drop in the FM electrode of the injector side (Co1 in this study) cannot be ignored, and position dependence of the electric field in the graphene channel takes place. The position dependence is maintained in the graphene channel, and induces position dependence of chemical potential in Co2 of the detector side, which is observed as the offset voltage. Actually, when we change the polarity of the electric current the sign of the offset also changes, and in addition, we have observed the fact that the offset voltage increased as the injection current increased. These observations collaborates the fact that the offset is attributed to the voltage drop in the FM electrode. The observed resistance hysteresis was not a spurious signal, but due to spin injection into the MLG, because we carefully fabricated the samples as the graphene channel was completely covered by the Co electrodes (as shown in Fig. 1) in order to avoid a local Hall effect, which can occur if the ferromagnetic electrodes do not cover the graphene channel completely. Hence, an MR effect through an unexpected electric pad-to-pad coupling cannot be observed.

Good correspondence of the non-local output voltage in the "cross" and "half" alignments[22] was verified (see Fig. 2(c)). When we assume that all contact resistances have same values for simplicity (see the discussion later), the non-local output voltage to be expressed as the following generalized form[23],

$$\Delta V_{non-local} = \frac{2P^2}{(1-P^2)^2}(\frac{R_F}{R_N})R_F \cdot [\sinh(\frac{L}{\lambda_{sf}})]^{-1} \cdot I_{inject}, \quad (1)$$

where P is the spin polarization, $\lambda_{sf}$ is the spin flip length, L is the gap length between two FM electrodes, $I_{inject}$ is the injected electric current, $R_F$ and $R_N$ are spin accumulation resistances of FM and NM, respectively, which is defined as (the conductivity)×(the spin diffusion length)/(the cross-sectional area). From equation (1), it is interpreted that the linear dependence is induced by constant spin polarization, and this finding manifested the robustness of the spin polarization at Co/MLG within ± 1 mA. Hanle-type spin precession experiments were implemented in order to verify that the obtained signals are ascribed to spin injection into the MLG. (Fig. 2(d)). For this purpose, we applied a magnetic field in the y direction in order to prepare the Co electrodes in a parallel or anti-parallel magnetization direction, and the field was then removed and a magnetic field, B, was scanned in the z direction (see Fig 1). Although raw data of spin precession was affected by magnetic-field-dependent background (not shown here) as already observed in GaAs spin valves[24], the clear crossing of parallel and anti-parallel signals was observed at +170 and -170 mT, which directly indicates that the observed signals were attributed to the spin injection into the MLG. By using the following equation for spin precession[11,25],

$$\frac{V_{non-local}}{I_{inject}} = \frac{P^2}{\sigma_{MLG} A/D} \int_0^\infty \frac{1}{\sqrt{4\pi Dt}} \exp(-\frac{L^2}{4Dt}) \cos(\omega_L t) \exp(-\frac{t}{\tau_{sf}}) dt, \quad (2)$$

where t is time, $\tau_{sf}$ is the spin coherent time and D is a diffusion constant, the spin diffusion constant D (=$2.1 \times 10^{-2}$ m²/s), spin flip length $\lambda_{sf}$ (=1.6 μm), spin coherent time $\tau_{sf}$ (=120 ps), and spin polarization P (=0.09) was estimated, although it should be noted that this is an approximate estimation because the Eq. (2) is the only existing expression for the estimation of spin transport properties in the case of the Hanle type spin precession.

The next point of interest is observation of the same robustness in the local scheme. Positive MR signals were detected between 14.5 and 18.5 mT, whereas negative MR signals were observed below 14.5 mT using a positive sweep with an external magnetic field (Fig. 3(a)). As pointed out in our previous letter[10], the negative MR is attributed to the existence of a magnetic domain wall during magnetization reversal in the Co2 electrode with a weaker coercive force, namely, due to the anisotropic MR (AMR) effect. The AMR effect, due to magnetization reversal of the Co1 electrode with a larger coercive force, is also seen as the gradual decrease of the positive MR at around 18.5 mT. Observation of AMR indicates that the measurement is sufficiently precise to detect such faint signals. The positive MR itself is not due to AMR, but is due to the spin injection, because the region of the magnetic field where the resistance hysteresis appears in the non-local and the local measurements coincides completely, that is, the spin injection into the MLG at RT was also clarified by the local scheme. The achievement of observing the local MR signals in molecular materials, where correspondence between "non-local" and "local" signals was clarified and the "local" signal was definitely due to spin accumulation at RT in graphene-based spin devices, has not been reported[26].

The injection current dependence of the output voltage is shown in Fig. 3(b), where the linear dependence, as observed in the non-local signal, is again seen here. From these results, we can elucidate that robustness of the spin polarization at Co/graphene within ±1 mA appears in the non-local and local signals. Here, the sample resistance was ~200 Ω, in which the resistance of one Co electrode wire and the MLG was measured to be ~50 Ω and ~5 Ω, respectively. This indicates that additional resistance (~50 Ω each) exists at a Co/graphene interface although no tunneling barrier such as Al-O was introduced. As a result, the spin polarization (=MR ratio) of this sample was constant up to ~100 mV at RT, which is surprising compared with the results of other molecular spin valves, such as a SWNT-spin valve, a MLG-spin valve and a rubrene TMR device, where no such robustness was observed. In order to determine the maximum voltage where the spin polarization is constant, another MLG device ($R_{Au-Co}$~109 Ω) was prepared and investigated using the non-local method, where the sample resistance without the Co wire resistance was ~60 Ω. Although the electrode (not the MLG channel) was broken at 20.3 mA, the current dependence of the output voltage exhibited very unique behavior (Fig. 4(a)). The output voltage exhibited the linear dependence (robustness of the spin polarization) until 0.5 V; above 0.5 V, it exhibited sub-linear dependence. However, even at ~1.2 V, the spin polarization was still 81% of the initial value (Fig. 4(b)). In addition, further experiments using the other samples exhibited that the robustness was maintained up to +2.7 V under a positive bias voltage application and down to -0.6 V under a negative voltage application (Fig. 4(c) and (d)). Here, it is noted that the results shown in Fig. 4(c) and (d) does not mean the robustness is asymmetric because the results were obtained in different samples and the bias voltage was applied until the samples were broken. Such robustness of spin polarization has never been expected and has not been observed in the other spintronics devices; therefore, the MLG device has exhibited the best performance with regard to robustness of the spin polarization (MR ratio) with respect to the bias voltage. The bias voltage dependence of the MR ratio (=the spin polarization of the injected spins) in spin devices is one of the most important issues which should be solved in order to spread application possibilities in a field of spintronics as mentioned above. Note that large MR ratio and robustness of spin polarization are independent physical aspects and both can be achieved simultaneously. The MR value itself is not essential for spin transport phenomena and there are several potential manners for controlling the MR ratio by which the value can be easily increased. For instance, shortening the gap length between two FM electrodes is the easiest manner because the MR ratio has exponential dependence to the gap length. An introduction of half metallic materials with 100% spin polarization is also one of the most appropriate manners and we can theoretically expect unlimited number of the MR ratio under an application of a bias voltage within a minority spin gap. Hence, our finding provides an important breakthrough to solve this currently important problem in a field of spintronics, and in addition, this study has shown an example that the superiority of nano-carbonaceous spintronics is experimentally manifested.

Several explanations for the decrease of spin polarization under the bias voltage applications have been so far proposed: (1) leakage path formation or two-step-tunneling by non-uniformity of a tunnel barrier, (2) magnon or phonon excitation or impurity scattering in a barrier or at an interface between the FM and a barrier, and (3) the density of states in the FM, that is, they are fundamental factors for the decrease. Our result clearly excludes the possibility of (3), because conventional Co electrodes were employed. In comparison with the non-local and local output voltage in the MLG spin valve, the ratio of both voltages is almost 2 as theory predicts (Fig. 4(e)), which directly indicates the interface of FM and MLG is ideally formed as realized in the case of metallic spin valves[18]. In the case of the MgO-TMR, it is believed that magnon/phonon excitation may be suppressed in a certain extent because of fewer defects at the interface between MgO and FM and thus the comparably larger $V_{half}$ was obtained. Analogous to the MgO case, the formation of an ideal interface in this study excludes the possibility of detrimental origins such as a leakage path, magnon/phonon excitation and non-uniformity at the interface, and thus the decrease of the spin polarization of the injected spins (= spin scattering) is strongly suppressed. An investigation concerning the disappearance of the linear bias voltage dependence can also corroborate the formation of the ideal interface. Figure 4(f) shows the bias voltage dependence of the spin injection signals (an upper panel) and that of 2-terminal resistance of the sample (a lower panel). The spin polarization of the injected spins was constant below 1 V, because the spin injection signals exhibited the linear bias voltage dependence. The 2-terminal resistance was observed to be constant in the same bias voltage region. However, the resistance increased when the bias voltage, $V_{bias}$, was larger than 1 V, where the linear bias voltage dependence disappeared. Once the linear dependence disappeared in the high bias voltage region ($V_{bias} > 1$ V), we observed drastic decrease of the intensity of the spin injection signals in the low bias voltage region ($V_{bias} < 1$ V) where the linearity had been observed, and also found that the linearity was not able to be seen any more. This finding can indicate that the degradation at the interface between the Co and the graphene probably induced the disappearance of the linearity, because a dominant part where the bias voltage was applied in the sample was the interface as discussed above.

Here, more important physical aspect is obtained from the result, namely, the good accordance of the theory and the experiment in the spin injection signals. Instead of the ratio of 2 in local and non-local schemes, large discrepancy between local and non-local results has been reported in the SWNT and the SLG spin valves[11,17], where the ratio was ~35 and ~10 in the SWNT and the SLG spin valves even at low temperatures, respectively. Our finding provides good accordance of the theory and the experiment in molecular spintronics at RT, which enables us to construct a steadfast physical basis in molecular spintronics as that in metallic spintronics.

## 3. Conclusive remarks

We implemented careful characterizations of spin transport properties in MLG spin valves, and found that the unprecedented robustness of the spin polarization of the injected spins into MLG. The spin polarization of the injected spins were constant up to +2.7 V and -0.6 V in positive and negative bias voltage applications, which is superior to all spintronics devices and also exhibits a part of superiority of molecular spintronics using graphene, although spin coherence, spin lifetime and output voltage due to spin injection should be improved as a next important milestone. This robustness was attributed to suppression of spin scattering due to the ideal interface formation. We also found a more important physical aspect, that is to say, the good accordance of the theory and the experiment, which was a missing part for constructing the steadfast basis of physics in molecular spintronics for further discussion of underlying physics.

**Experimental detail**

The starting materials used for preparation of the MLG spin valves were highly oriented pyrolytic graphite (HOPG, NT-MDT Co.) and polyimide-oriented highly oriented graphite (Super graphite, Kaneka Co.)[27]. MLG flakes were peeled from these materials using adhesive tape. The flakes were then pushed onto the surface of a $SiO_2$/Si substrate ($SiO_2$ thickness = 300 nm). The typical thickness of MLG that provided observable spin injection signals in a spin valve structure was 2-40 nm. The non-magnetic and ferromagnetic electrodes used were Au/Cr (=40/5 nm) and Co (=50 nm), respectively, and were patterned using an electron beam lithography technique. The width of the Co electrodes, Co1 and Co2, were the same, but Co2 possessed a pad structure in order to weaken the coercive force, and the gap width of the Co electrodes was 1.5 μm (Fig. 1). All measurements of MR effects were performed at RT. We introduced a non-local scheme[25] in addition to a conventional local scheme for excluding spurious signals. One can detect a non-local output voltage which is induced by position dependence of electrochemical potential of the generated spin current in the MLG. Here, spins are injected at ferromagnet(FM)/graphene of the top layer of the MLG in the injector side (between Co1 and Au1) as an electric current and accumulated spins diffuse from Co1 to Co2 and Au2 (the detector side). Hence, an output voltage induced by the generated spin current is determined by an amount of accumulated spins, namely, it is strongly affected by the interface spin polarization at Co1/graphene. In a non-local scheme, an electric current I is injected from Co1 into GTF and extracted at Au1. The voltage difference is measured between Co2 and Au2. The non-local voltage, $V_{non-local}$, is defined as $(V_+ - V_-)$. In a local scheme, an electric current I is injected from Co1 into GTF and extracted at Co2. The voltage difference is measured also between Co1 and Co2. The "local" voltage, $V_{local}$, is defined as $(V_+ - V_-)$. The spin injection was investigated using a four terminal probe system (ST-500, Janis Research Company Inc.) with an electromagnet. The magnetic field was swept from -400 Oe to +400 Oe in steps of 8 Oe. A source-meter (Keithley Instruments Inc., KH2400) and a multi-meter (Keithley Instruments Inc., KH2010) were used to detect spin injection signals. The Hanle effect was investigated using a physical property measurement system (PPMS, Quantum Design Inc.) at RT. The magnetic field was swept from -200 mT to +200 mT in ca. 3 mT steps. The initial magnetization configuration of Co1 and Co2 was set to be either parallel or anti-parallel.

## Acknowledgments

The authors would like to thank Dr. M. Murakami and Prof. K. Shimizu for providing source materials for the MLG, Dr. T. Seki for fruitful discussion, and S. Tanabe, K. Muramoto and T. Sugimura for experimental assistance. M.S. thanks Prof. T. Takenobu for his critical reading of the manuscript, and thanks G. Niebler for providing graphical arts of spin transport in graphene.

**Figures and Figure captions**

**Figure 1 (color online)**| Optical microscopic image of a four-terminal MLG spin valve.

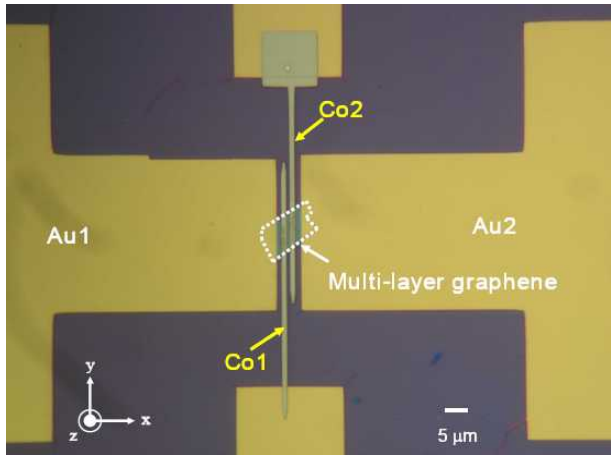

**Figure 2 (color online)|** (a) Schematic cross sectional view of the MLG spin valve, the set-up for the non-local measurement and the injection current dependence of the non-local resistance ($V_{non-local}/I_{inject}$). The injection current was swept from -1 mA to +1 mA. The value of the non-local MR was always observed to be approximately 18 mΩ. (b) Injection current dependence of the output voltage, $V_P$-$V_{AP}$. Linear dependence, within ±1 mA, is clearly exhibited. (c) Spin injection signals of the "half" and the "cross" alignments, respectively, for an injection current of 800 μA. The output voltages in both cases were 5 μV and were in correspondent. (d) Modulation of the non-local resistance due to spin precession, as a function of perpendicular magnetic field. The injection current was set at 300 μA. The red (parallel, ↑↑) and blue (anti-parallel, ↑↓) open circles are experimental data, and the red and blue lines are the result of model fitting.

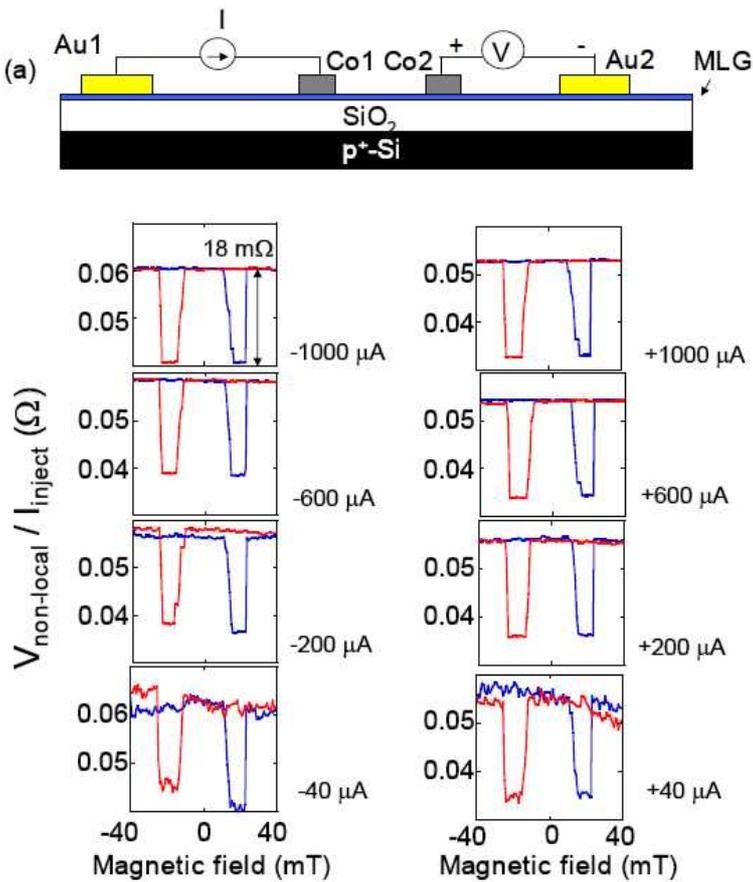

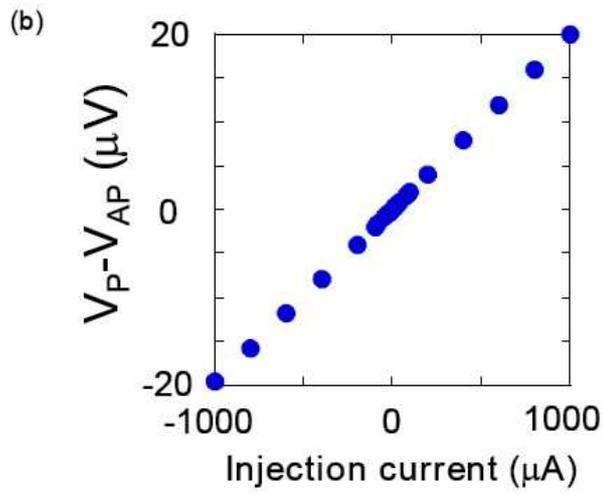

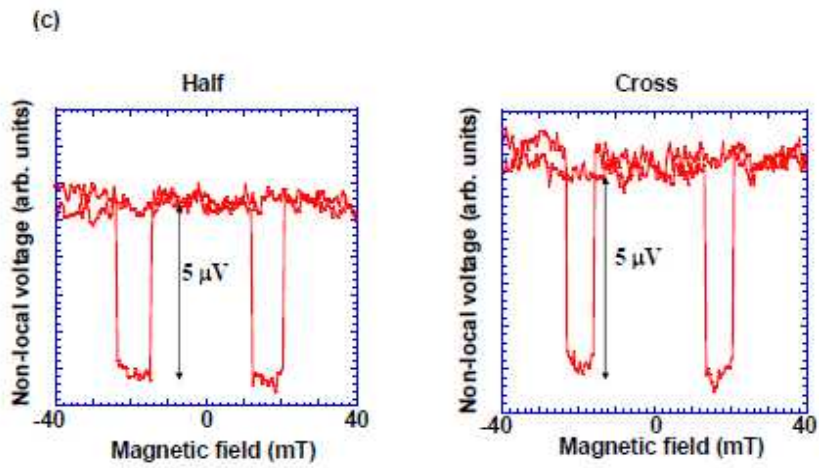

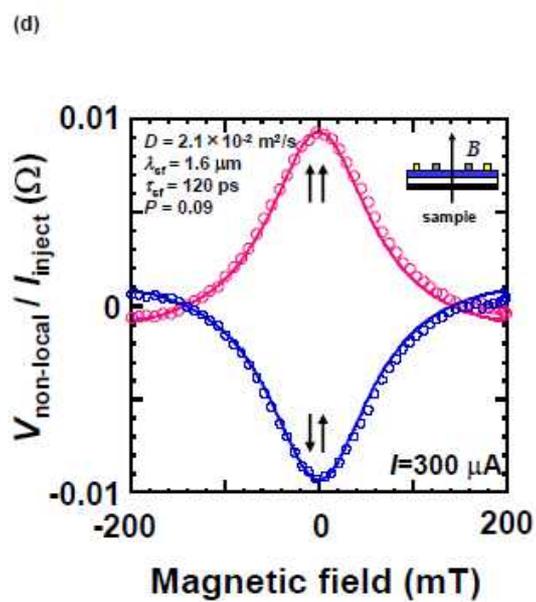

**Figure 3 (color online)|** (a) Cross sectional schematic of the set-up for the local measurement and injection current dependence of the local resistance ($V_{local}/I_{inject}$). The injection current was swept from -1 mA to +1 mA. The MR ratio was defined as $\{(R_{anti\text{-}parallel} - R_{parallel})/R_{parallel}\}$ and was always observed to be approximately 0.02%. The data of $I_{inject}$=40 µA includes some noise, and the spin injection signals may not be clear enough; however, these data signals were not anomalously large and linear dependence was thought to be maintained. (b) Injection current dependence of the local output voltage ($V_{AP}$-$V_P$), which exhibits linear dependence within $\pm 1$ mA.

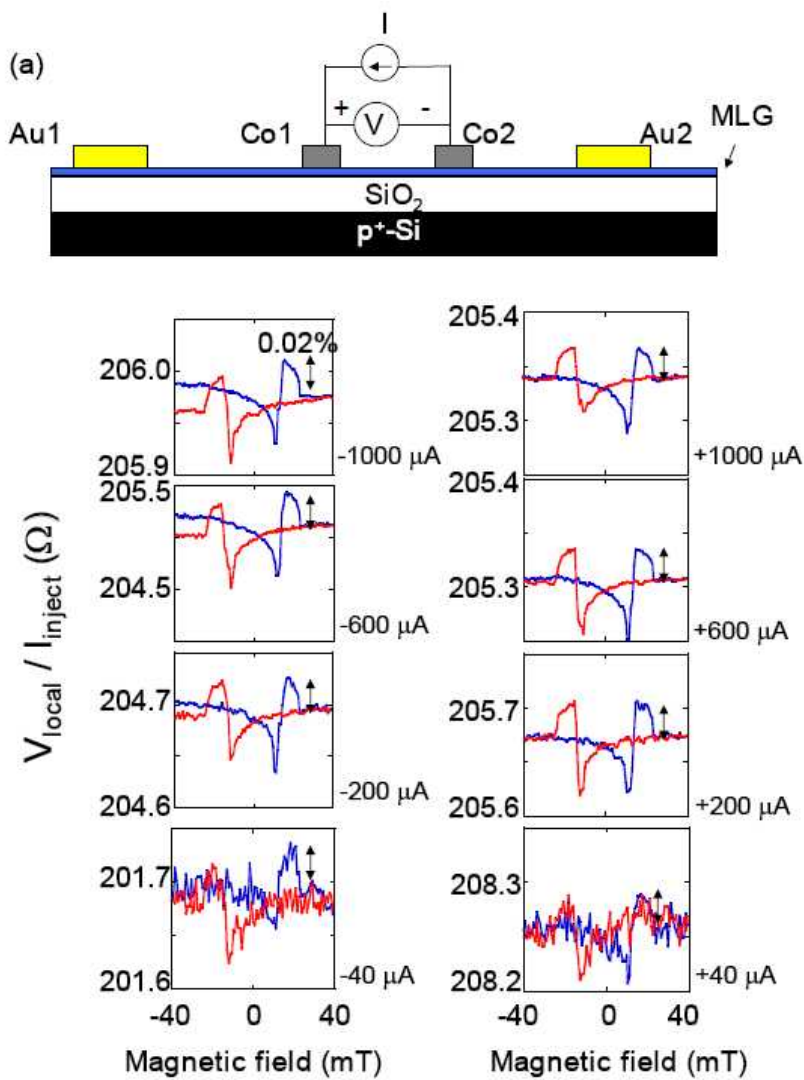

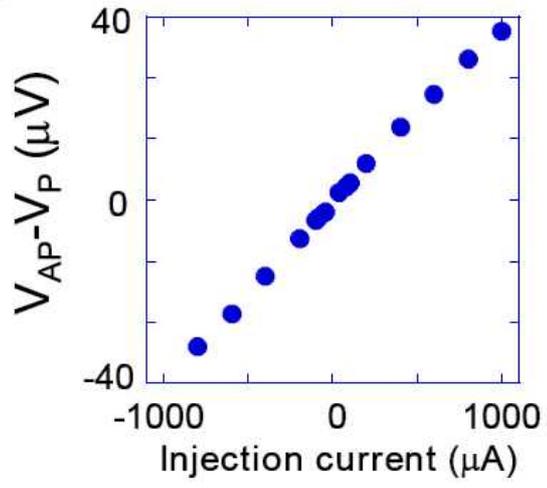

**Figure 4 (color online)**| **(a)** Bias voltage dependence of the non-local signals. The dashed line is the line fitted under the assumption that linear dependence is maintained until 10 mA. **(b)** Bias voltage dependence of the spin polarization in a MgO-based TMR device (red dashed line[6]) and the MLG device of this study (blue solid line). Normalized spin polarization in the MLG device was estimated by calculating the ratio of the experimentally obtained values and the values on the fitted line under the assumption of linear dependence (Fig. 4(a)). **(c)(d)** Results on the robustness experiment in the other sample in positively biased ((c)) and negatively biased ((d)) conditions. The spin polarization is constant up to +2.7 V under a positive bias voltage application, and is constant down to -0.6 V. **(e)** Comparison of the values of the spin injection signals in the MLG spin valve for the non-local and the local methods. The polarity of the signal in the non-local method was reversed, and the non-local and the local resistance was normalized for ease of understanding. **(f)** Correspondence between 2-terminal resistance of the sample and the spin injection signals. The upper panel shows the bias voltage dependence of the spin injection signals, and the numbers in the panel indicates the 2-terminal sample resistance. The lower panel shows the relationship between the sample resistance and the bias voltage. The dashed lines in the panels are guide for the eye.

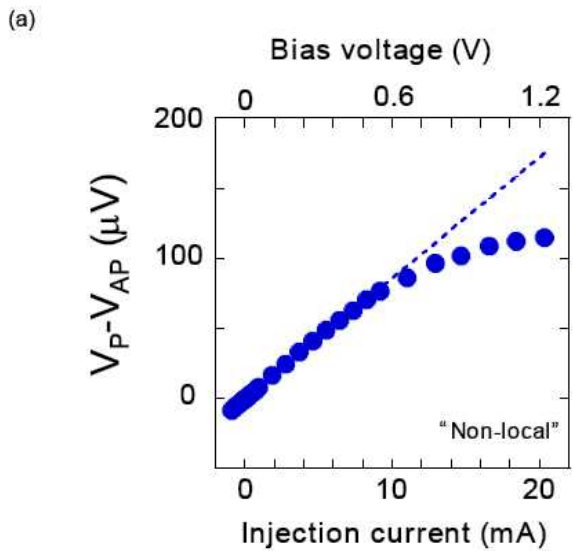

(b)

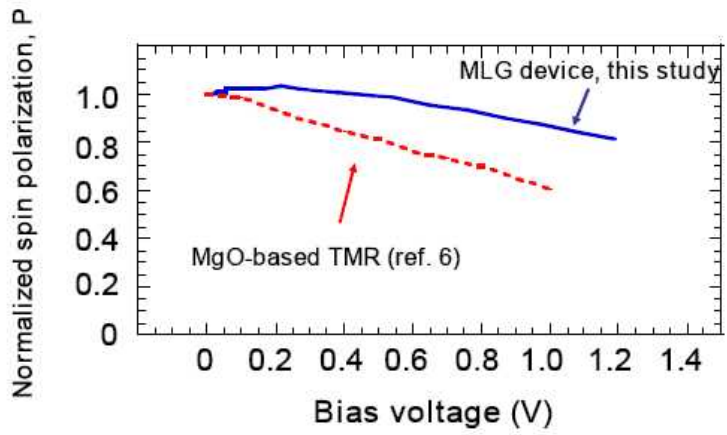

(c)

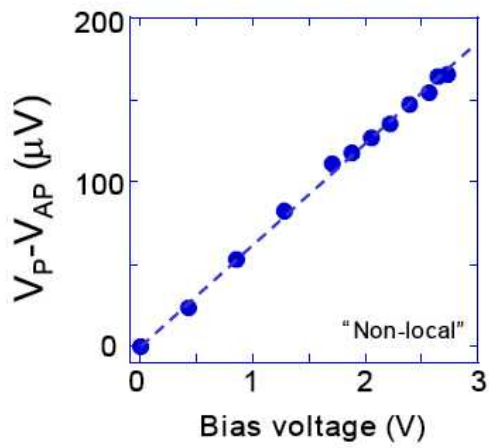

(d)

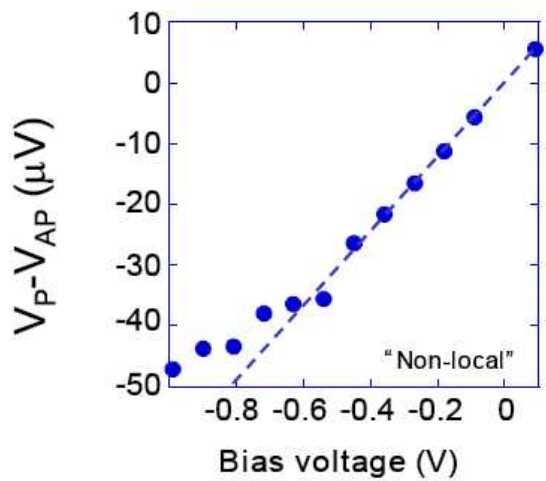

(e)

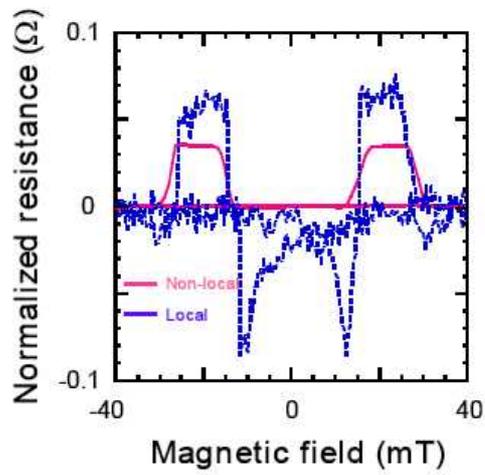

(f)

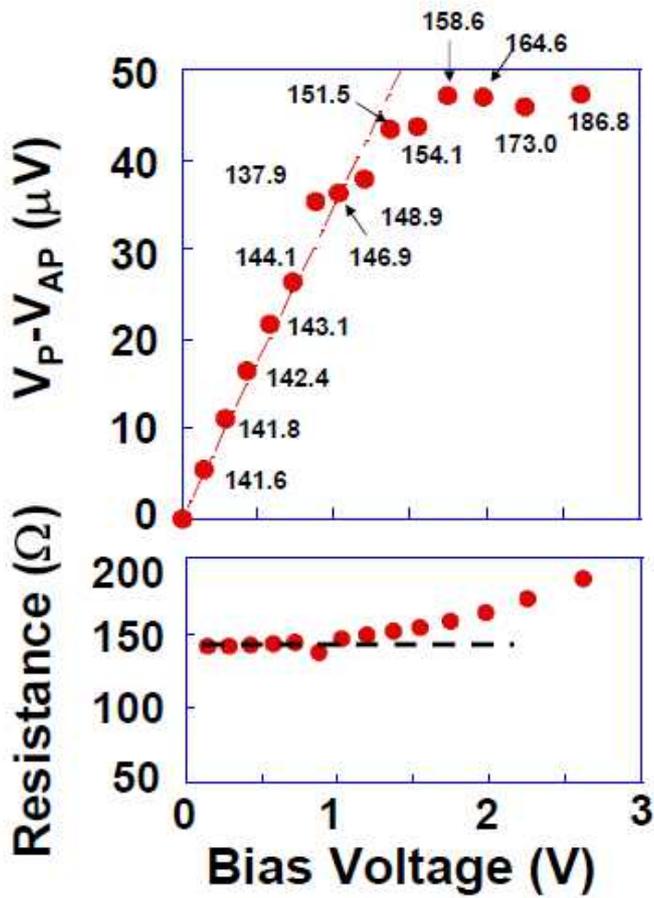